\documentstyle[prc,preprint,tighten,aps]{revtex} 
\begin{document}
\draft
\preprint{\vbox{\noindent 
 \hfill LA-UR-98-9\\
 \null\hfill nucl-th/9802003}}
\title{Effects of $\bbox{^8}$B size  
on the low-energy
$\bbox{^7}$Be($\bbox{p,\gamma}$)$\bbox{^8}$B cross
section} 
\author{Attila Cs\'ot\'o$^{a,b,c}$ and Karlheinz Langanke$^b$}
\address{$^a$Theoretical Division, Los Alamos National 
Laboratory, Los Alamos, NM 87545, USA\\
$^b$Institute for Physics and Astronomy and Center for 
Theoretical Astrophysics, Aarhus University, DK-8000 Aarhus, 
Denmark\\
$^c$Department of Atomic Physics, Eotvos University, Puskin utca 5-7,
H-1088 Budapest, Hungary } 
\date{\today}

\maketitle

\begin{abstract}
\noindent
We calculate several ``size-like'' $^8$B observables within the 
microscopic three-cluster model and study their potential 
constraints on the zero-energy astrophysical $S_{17}(0)$ factor of 
the  $^7{\rm Be}(p,\gamma){^8{\rm B}}$ reaction. We find 
within our three-cluster model that a simultaneous reproduction of
the experimental data for the
$^8$B radius and quadrupole moment and of the $^8$B-$^8$Li Coulomb
displacement energy implies
$S_{17}(0)=(23-25)$ eV$\,$b. 
\end{abstract}

\narrowtext

\section{Motivation}

The $^7{\rm Be}(p,\gamma){^8{\rm B}}$ reaction is currently 
considered to be one of the astrophysically most important 
nuclear reactions, as its low-energy cross section 
determines the high-energy solar neutrino flux 
\cite{Bahcall}. Recently there has been a great deal of 
experimental and theoretical activities investigating this 
process. The low-energy cross section has been studied 
directly by using a radioactive $^7$Be target and a proton 
beam \cite{Hammache}, in inverse kinematics by using a 
$^7$Be beam and a proton target \cite{Campajola},  
indirectly from the Coulomb dissociation of $^8$B 
\cite{Motobayashi}, and by extracting the $^7{\rm Be}+p$ 
nuclear vertex constant from the $^7{\rm Be}(d,n){^8{\rm 
B}}$ reaction \cite{Liu}. On the theoretical side some
effects of $^7$Be deformations \cite{Nunes} and three-body
dynamics \cite{Grigorenko} have been studied, and efforts 
to understand the nuclear vertex constant have been made
\cite{Timofeyuk}. 
In Ref. \cite{PRC} we have shown that the zero-energy cross section
of the $^7{\rm Be}(p,\gamma){^8{\rm B}}$ reaction
scales linearly with the, unfortunately yet unknown, quadrupole moment
of $^7$Be. In the present paper we extend this study and investigate
the relation
between the zero-energy cross section and several $^8$B ``size''
properties.

In our approach we study the ground state properties of 
$^7$Be and $^8$B as well as the $^7{\rm Be}(p,\gamma){^8{\rm 
B}}$ reaction cross section consistently within 
the microscopic eight-body $^4{\rm He}+{^3{\rm He}}+p$
cluster model. As this model has been discussed before
we refer the reader to  Refs.\ \cite{Csoto94,PRC} for 
details of the theoretical background. As customary in 
nuclear astrophysics we define the cross sections in terms 
of the astrophysical $S$ factor
\begin{equation}
S(E)=\sigma (E)E\exp{\Big [2\pi\eta (E)\Big ]}, \hskip 0.5cm
\eta (E)={{Z_1Z_2e^2}\over{\hbar v}} ,
\end{equation}
where $Z_1,Z_2$ are the charges of the two colliding nuclei, 
and $v$ is their relative velocity.

At low, and in particular, at solar energies the $^7{\rm 
Be}(p,\gamma){^8{\rm B}}$ reaction is highly peripheral, 
which means that only the external parts of the bound- and
scattering wave functions contribute to the radiative 
capture cross section \cite{Langanke}. The external wave functions are known 
with the exception of the asymptotic normalization constant, 
$\bar c$, of the $^8$B bound state \cite{Xu}. Consequently 
the energy dependence of the low-energy $S_{17}(E)$ factor 
is well-known (e.g. Refs. \cite{Williams,Langanke,Csoto97}). 
Its absolute value, however, depends on 
${\bar c}$ and has thus to be determined experimentally. 
Nevertheless theoretical constraints on the asymptotic 
normalization constant might be quite useful. We note that 
$\bar c$ depends mainly on the effective $^7$Be--$p$ 
interaction radius. A larger radius results in a lower 
Coulomb barrier, which leads to a higher tunneling 
probability into the external region, and hence to a higher 
cross section. A possible way to constrain the interaction  
radius is to study some key properties of the $A=7$ and 8 
nuclei \cite{ENS}. The observables that are most sensitive 
to the interaction radius are ``size-like''  properties, for 
example, quadrupole moment, radius, Coulomb displacement 
energy \cite{Brown}, etc. These are the quantities which we 
will calculate in our microscopic cluster model and then 
study their effect on the astrophysical $S$ factor.

\section{The size of $\bbox{^8}$B and its effect on 
S$\bbox{_{17}}$}

In Ref.\ \cite{PRC} we demonstrated that there is a 
linear correlation between the zero-energy astrophysical
$S$ factor, $S_{17}(0)$, of the $^7{\rm Be}(p,\gamma){^8{\rm 
B}}$ reaction and the quadrupole moment of $^7$Be, $Q_7$. 
The $Q_7$ quadrupole moment has not been measured yet, but 
in Ref. \cite{PRC} we have predicted it to be between $-6$ 
e$\,$fm$^2$ and $-7$ e$\,$fm$^2$. The absolute scale of the 
$S_{17}(0)-Q_7$ correlation, however, depends on the applied 
effective nucleon-nucleon ($N-N$) interaction. For our
preferred MN interaction this resulted in $S_{17}(0) = 25-26.5$ eV$\,$b 
\cite{PRC}. Other interactions gave slightly larger or smaller 
$S_{17}(0)$ values, but these interactions were found to be 
inferior to the MN force for other observables. Now we will 
turn to the other ``size-like'' observables and their 
potential constraints on $S_{17}(0)$. For this purpose we 
have repeated the microscopic calculations described in
Ref.\ \cite{PRC} varying the size parameter of the $^3$He 
and $^4$He clusters while keeping other important ingredients 
of the calculation fixed. Again as in Ref.\ \cite{PRC} we 
have performed these calculations for several interactions 
(MN force \cite{MN}, V2 interaction \cite{V2}, and MHN 
interaction \cite{MHN}). Our calculation thus reproduces the
$S_{17}(0)-Q_7$ plot shown in Fig.\ 1 of Ref.\ \cite{PRC}. 

In the present Fig. 1 we extend the study of Ref. \cite{PRC} and
investigate the relation of $S_{17}(0)$ to several size-like properties
of $^8$B:
(a) the $^8$B radius r($^8$B), (b) the difference between the 
$^7$Be and $^8$B radii quantified by $r^2({^8{\rm 
B}})-r^2({^7{\rm Be}})$,  (c) the $^8$B quadrupole moment, 
 and (d) the $E({^8{\rm Li}})-E({^8{\rm B}})$ Coulomb 
displacement energy. These calculations have been 
performed for the same model spaces and interactions as in 
Ref.\ \cite{PRC}. Importantly all four indicators scale linearly
with $S_{17}(0)$. This is caused by the `halo' structure of the
$^8$B ground state \cite{Riisager} and reflects that the
$^7$Be(p,$\gamma$)$^8$B reaction at low energies is an external capture
process. In the following we will discuss the four indicators in turn
and will try to derive at possible constraints on $S_{17}(0)$.
First, the comparison is performed for the consistent eight-body
$^4$He+$^3$He+p calculation using the MN force; the dependence of our
results on the model space and the interaction employed will be
discussed below.

The size property that is most sensitive to the effective 
$^7{\rm Be}-p$ interaction radius is $r^2({^8{\rm 
B}})-r^2({^7{\rm Be}})$. However, a precise experimental 
determination of this quantity is very difficult. In fact, 
during the course of the work reported in Ref.\ \cite{PRC} it seemed 
hopeless that the $^7$Be or $^8$B radius could be measured 
with relatively high precision. The radii of nuclei far 
from stability are usually extracted from interaction cross 
section measurements by using Glauber-type models with 
uniform density distribution for the nuclei \cite{Tanihata}. 
Recently, a new and  more precise method has been 
introduced  \cite{AlKhalili} which 
considers the few-body structure of the nuclei involved 
(like $^7{\rm Be}+p$ for $^8$B), while extracting
the radius from the measured interaction cross sections. For 
$^8$B the resulting point-nucleon radius is $r({^8{\rm 
B}})=2.50\pm0.04$ fm, and hence $r^2({^8{\rm 
B}})-r^2({^7{\rm Be}})\approx0.9$ fm$^2$. We note, that for 
$^7$Be the model of Ref.\ \cite{AlKhalili} still uses the 
Glauber estimate. Our calculation reported in Ref. \cite{PRC}
gives 
$r({^8{\rm B}})= 2.73$ fm and thus 
overestimates the experimental value.
As expected we observe that $S_{17}(0)$ 
decreases with decreasing $^8$B radius.  
Using the linear relationship between 
$r({^8{\rm B}})$ and $S_{17}(0)$ and the experimental value for the
$^8$B radius places the cross section
into the range $S_{17}(0)= 23.2-24.2$ eV b.
Ref. \cite{PRC} found
$r^2({^8{\rm B}})-r^2({^7{\rm Be}})=0.8$ fm$^2$, which appears to be a
reasonable value. However, due to the uncertainties in the
phenomenological value the derivation of a constraint on $S_{17}(0)$ from
$r^2({^8{\rm B}})-r^2({^7{\rm Be}})$ is 
currently not possible.

The experimental value of the $^8$B quadrupole moment is 
$Q_8=(6.83\pm0.21)$ e$\,$fm$^2$ \cite{Minamisono}. In Ref. \cite{PRC} we
calculated a slightly larger value, $Q_8 =7.45$ fm$^2$.
Like in the case of the $^8$B radius, $S_{17}(0)$ increases linearly with
the $^8$B quadrupole moment. From a comparison to the
experimental data we find the constraint $S_{17}(0)=23.7-24.8$ eV b (Fig.
1c).

To derive a phenomenological value for the 
Coulomb displacement energy to be compared with our calculated 
values, we have to consider that the physics that accounts 
for the Nolen-Schiffer effect \cite{NS} is not present in 
our model. This effect is estimated to cause a $\approx$130 
keV shift in the $E({^8{\rm Li}})-E({^8{\rm B}})$ Coulomb 
displacement \cite{Brown}. So we should compare our results 
to a phenomenological value of  
$\Delta=3.54-0.13=3.41$ MeV. 
Ref. \cite{PRC} found a too small Coulomb displacement energy, 
$\Delta=3.2$ MeV. From Fig. 1d we observe that $S_{17}(0)$ decreases
linearly with $\Delta$; thus the experimental value for the Coulomb
displacement energy corresponds to $S_{17}(0)=24.3$ eV b.

We can summarize
the results for the complete $^4$He+$^3$He+p cluster
study (with the MN force) of Ref.  \cite{PRC} as follows:
This calculation \cite{PRC} gives consistently 
values for those indicators, for which reliable experimental data exist,
which point to the use of a too large $^7$Be+p interaction radius. 
This implies that
the value for $S_{17}(0)$ (26.1 eV b) predicted in \cite{PRC} is too
large. We note that, by slightly varying the interaction radius, 
all three experimentally determined ``size-like'' parameters (the $^8$B
radius and quadrupole moment and the Coulomb displacement energy) can
consistently be reproduced (see Table 1); the corresponding value
for the    
$^7$Be(p,$\gamma$)$^8$B reaction cross section then is
$S_{17}(0)=(23-25)$ eV$\,$b. 

How much do our results depend on the chosen model space 
and the adopted interaction?
To answer these 
questions we have at first performed a series of restricted 
calculations involving only ($^3{\rm He}+{^4{\rm He}})+p$ configurations 
($^7{\rm Be}+p$ like configurations) rather than all possible 
arrangements of the three clusters. (As in all calculations 
reported in this paper the experimental value of the $^8$B 
binding energy relative to the $^7{\rm Be}+p$ threshold has been 
reproduced by a slight modification in the $N-N$ interaction, 
see Ref.\ \cite{PRC}.) Again we find that $S_{17}(0)$ scales linearly
with all indicators. From Fig.\ 1 we also observe that, for a 
fixed value of $S_{17}(0)$, the extension of the model space 
(going from the restricted space to the full three-cluster model) 
reduces the $^8$B radius slightly, but increases the $^8$B 
quadrupole moment and the Coulomb displacement energy. The 
reason why the radius (a) and the quadrupole moment (c) of 
$^8$B change in opposite direction if the model space is 
enlarged is that the addition of the $^5{\rm Li}+{^3{\rm
He}}$ and $^4{\rm He}+{^4{\rm Li}}$ channels brings in large
charge polarization which increases the quadrupole moment
even if $r({^8{\rm B}})$ is reduced. 
Even if we allow the variation of the $^7$Be+p interaction radius, the
restricted model space calculation does not simultaneously reproduce the
experimental data for our indicators (see Table 1). 
While the $^8$B radius and the
Coulomb displacement energy puts $S_{17}(0)$ at around $23-24$ eV b, the
$^8$B quadrupole moment favors a larger value of $S_{17}(0)=25.3-27.5$
eV b. In fact, the $^8$B quadrupole moment is the quantity which is
clearly most sensitive to the model space. Note that enlargening the
model space does not only reduce $S_{17}(0)$ for fixed value of $Q_8$,
it also changes the slope of the $S_{17}(0)-Q_8$ scaling.
Obviously the reproduction of the $^8$B quadrupole moment requires a
3-body approach and is sensitive to the internal structure of the
clusters.

In Ref. \cite{PRC} it has been observed that the $S_{17}(0)-Q_7$ scaling
depends on the adopted NN interaction. We have therefore repeated
the $^7{\rm
Be}+p$-type model calculations  with  two 
other interactions (V2 and MHN).
We observe that, for  
fixed values of the $^8$B radius, of the 
$r^2({^8{\rm B}})-r^2({^7{\rm Be}})$ difference and of
the Coulomb displacement, the rather repulsive V2 interaction gives 
larger $S_{17}(0)$ values  than the MN 
interaction, while the MHN interaction gives smaller values. 
Assuming a linear relation betwen $S_{17}(0)$ and our indicators,
constraints on the zero-energy S-factor can be derived; the
corresponding values are listed in Table 1. For the MHN interaction we
find that the $^8$B quadrupole moment and the other indicators are not
simultaneously reproduced in the restricted model space. For the V2
interaction the $^8$B quadrupole moment and radius and the Coulomb
displacement energy are reproduced for a $^7$Be+p interaction radius
corresponding to $S_{17}(0) \approx 26$ eV b. 
However, for this value of $S_{17}(0)$ the
$r^2({^8{\rm B}})-r^2({^7{\rm Be}})$ difference becomes
unreasonably large. 

We note again that a measurement of the $^7$Be quadrupole moment would
place some additional constraints on the consistency of our
calculations. For the complete $^4$He+$^3$He+p model calculation the
simultaneous reproduction of the indicators predict $Q_7$ to be in the
range $-(5.5-6.0)$ e fm$^2$. 
However, this value is smaller than the one
($Q_7=-6.9$ e$\,$fm$^2$ \cite{PRC}) obtained if we chose the 
cluster size parameters such to reproduce the quadrupole 
moment of the analog nucleus $^7$Li.
Does this already point to the 
necessity of a further enlargement of the model space
beyond the $^4{\rm He}+{^3{\rm He}}+p$ three-cluster model 
which  would then also effect our results obtained for $^7$Be, 
e.g., change the $^7$Be quadrupole moment? To investigate 
this we have performed calculations for $^7$Be in which we 
have added the $^6{\rm Li}+p={^4{\rm He}}+d+p$ configuration 
to the $^4{\rm He}+{^3{\rm He}}$ model space, adopted above. 
In all cases the exchange mixture parameter of the $N-N$ 
interaction was fixed to reproduce the $^7$Be binding energy 
relative to the $^4{\rm He}+{^3{\rm He}}$ threshold. We also 
made sure that the $^6{\rm Li}+p$ threshold was correctly 
reproduced. Our results show that $\vert Q_7\vert$ is 
increased by $0.5-1$ e$\,$fm$^2$ in the coupled-channel 
model relative to the single-channel $^4{\rm He}+{^3{\rm
He}}$ value perhaps suggesting the need for an even larger model space than the
complete $^4$He+$^3$He+p model space adopted here. Which consequences
such an enlargement might have on $S_{17}(0)$ has to wait for 
$^8$B calculations
performed in larger model spaces, which are beyond $^4{\rm
He}+{^3{\rm He}}+p$.

\section{Summary}

In summary, we have adopted a microscopic $^4{\rm
He}+{^3{\rm He}}+p$ cluster model to calculate the $^8$B 
radius and quadrupole moment, the difference in the $^7$Be and 
$^8$B radii, $r^2({^8{\rm B}})-r^2({^7{\rm Be}})$, and the 
Coulomb displacement energy $E({^8{\rm Li}})-E({^8{\rm B}})$ 
and to study their relation to the $S_{17}(0)$ astrophysical 
$S$ factor. We find that all these indicators scale linearly with the
zero-energy $^7$Be(p,$\gamma$)$^8$B cross section. 

Within our three-cluster model we find that the 
experimentally determined values for the $^8$B radius and 
quadrupole moment as well as the Coulomb displacement energy 
is consistently described  if the internal cluster size 
parameters are chosen such that the values for  
$S_{17}(0)$ are  between ($23-25$) 
eV$\,$b. This range is  more or less compatible with the value 
currently used in most solar models, $S_{17}(0)=22.4\pm2.1$ 
eV$\,$b \cite{Johnson}. However, it is slightly inconsistent 
with the recently adopted new experimental value 
$S_{17}(0)=19^{+4}_{-2}$ eV$\,$b \cite{Adelberger}. 
 
As a note of caution we mention that this result has been 
derived from a linear relation between our four indicators
and $S_{17}(0)$ 
found in our three-cluster model. However, we found that 
enlarging the $^7$Be model space by adding a $^4{\rm 
He}+d+p$ configuration to the $^4{\rm He}+{^3{\rm He}}$ 
configuration increased the $^7$Be quadrupole moment by 
about $10 \%$. To investigate the effects which additional 
configuration might have on the $^8$B properties and in
particular on the $S_{17}(0)$ value, requires calculations 
in model spaces which are beyond $^4{\rm He}+{^3{\rm
He}}+p$. 

\mbox{\ }

The work of A.\ C.\ was performed under the auspices of the 
U.S.\ Department of Energy. The work has been partly supported by the
Danish Research Council and by OTKA grant F019701.

\begin{table}
\caption
{The constraints derived for $S_{17}(0)$ (in eV b)
from the $^8$B radius and quadrupole moment and from the Coulomb
displacement energy
in our complete $^4$He+$^3$He+p
8-body calculation (full) and in the restricted $^7$Be+p model spaces
for the Minnesota force (MN), the Volkov force (V2) and the
Hasegawa-Nagata force (MHN).}
\begin{tabular}{c|c|c|c|c}
indicator &full & MN &  V2 & MHN \\ \hline
r($^8$B) & $23.2-24.2$ & $22.8-23.6$ & $25.7-26.6$ & $21.7-22.7$ \\
$Q_8$    & $23.7-24.8$ & $25.3-27.5$ & $24.1-27.0$ & $24.3-27.2$ \\
$\Delta$ &  24.3       &   23.8      &   26.5     &    23.0 \\   
\end{tabular}
\end{table}

\widetext
\begin{figure}
\caption{Correlation between the zero-energy astrophysical 
$S$ factor of the $^7{\rm Be}(p,\gamma){^8{\rm B}}$ reaction 
and (a) the $^8$B point-nucleon radius, $r({^8{\rm
B}})$ (in fm), (b) the $r^2({^8{\rm B}})-r^2({^7{\rm Be}})$ 
value (in fm$^2$), (c) the quadrupole moment of $^8$B, $Q_8$ 
(in e$\,$fm$^2$), and (d) the $\Delta=E({^8{\rm Li}})-E({^8{
\rm B}})$ Coulomb displacement energy (in MeV). The correlations have
been calculated
in our microscopic 
eight-body model,  using 
several  $N-N$ interactions and model spaces. 
For a given model space and interaction, different results
are obtained by varying the cluster size parameters. For a 
detailed description of the model spaces and interactions, 
see Ref.\ \protect\cite{PRC}. 
The 
phenomenological values are $r({^8{\rm B}})=2.50\pm0.04$ fm 
\protect\cite{AlKhalili}, 
$r^2({^8{\rm B}})-r^2({^7{\rm Be}})\approx0.9$ fm$^2$, 
$Q_8=6.83\pm0.21$ e$\,$fm$^2$ \protect\cite{Minamisono}, 
and $\Delta=3.41$ MeV. In the Coulomb displacement energy the
Nolen-Schiffer anomaly is removed from the value given in
Ref.\ \protect\cite{Ajzenberg} (see the text). The phenomenological
values are
indicated by dashed lines in the respective figures.}
\label{fig1}
\end{figure}

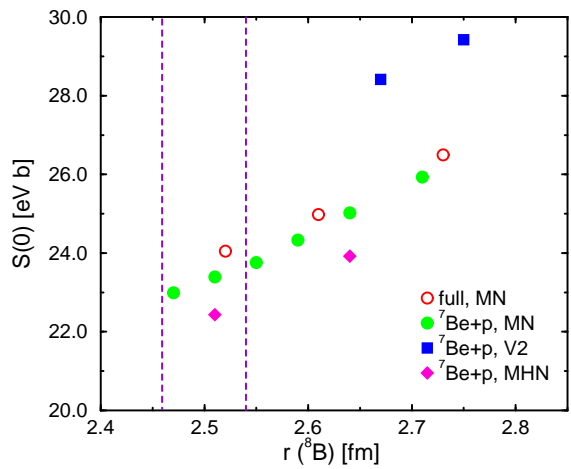
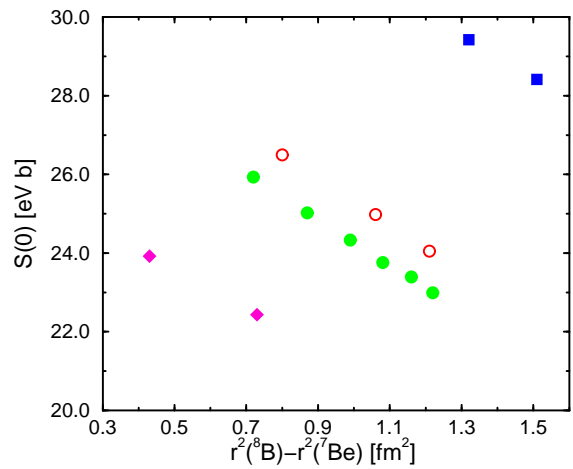
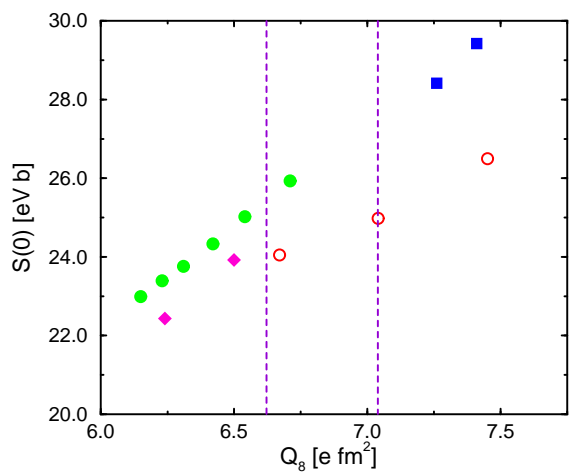
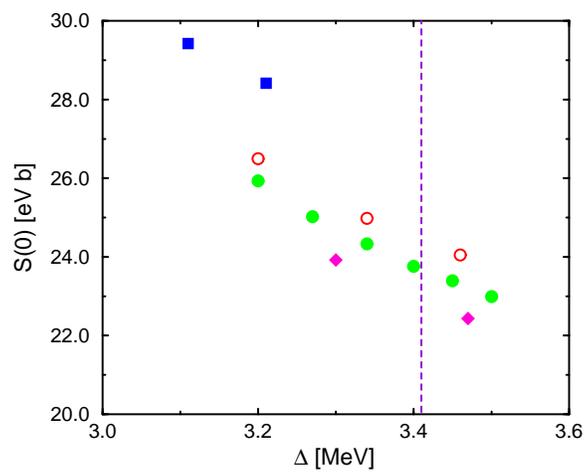
\end{document}